\documentclass[prd,onecolumn,notitlepage,eqsecnum,nofootinbib,floatfix]{revtex4-1}
\usepackage{amssymb}
\usepackage{amsmath}
\usepackage{amsfonts} 
\usepackage{bm}
\newcommand{\E}{{\cal E}} 
\newcommand{\B}{{\cal B}} 
\newcommand{\K}{{\cal K}} 
\newcommand{\Bq}{{\cal B}^{\scriptstyle \sf q}} 
\newcommand{\Kd}{{\cal K}^{\scriptstyle \sf d}} 
\newcommand{\Bhatq}{\hat{\cal B}^{\scriptstyle \sf q}} 
\newcommand{\Ko}{{\cal K}^{\scriptstyle \sf o}} 
\newcommand{\m}{{\sf m}} 
\newcommand{\bq}{b^{\scriptstyle \sf q}} 
\newcommand{\bhatq}{\hat{b}^{\scriptstyle \sf q}} 
\newcommand{\kd}{k^{\scriptstyle \sf d}} 
\newcommand{\ko}{k^{\scriptstyle \sf o}} 
\newcommand{\xid}{\xi^{\scriptstyle \sf d}} 
\newcommand{\xio}{\xi^{\scriptstyle \sf o}} 
\newcommand{\xiq}{\xi^{\scriptstyle \sf q}} 
\newcommand{\xihatq}{\hat{\xi}^{\scriptstyle \sf q}} 
\newcommand{\vd}{v^{\scriptstyle \sf d}} 
\newcommand{\vo}{v^{\scriptstyle \sf o}} 
\newcommand{\vq}{v^{\scriptstyle \sf q}} 
\newcommand{\vhatq}{\hat{v}^{\scriptstyle \sf q}} 
\newcommand{\pd}{p^{\scriptstyle \sf d}} 
\newcommand{\po}{p^{\scriptstyle \sf o}}

\allowdisplaybreaks
\begin{document}
\title{Dynamical response to a stationary tidal field}   
\author{Philippe Landry} 
\affiliation{Department of Physics, University of Guelph, Guelph,
  Ontario, N1G 2W1, Canada} 
\author{Eric Poisson} 
\affiliation{Department of Physics, University of Guelph, Guelph,
  Ontario, N1G 2W1, Canada} 
\date{October 30, 2015} 
\begin{abstract} 
We demonstrate that a slowly rotating compact body subjected to a
stationary tidal field undergoes a dynamical response, in which the
fluid variables and the interior metric vary on the time scale of the
rotation period. This dynamical response requires the tidal field to  
have a gravitomagnetic component generated by external mass currents; 
the response to a gravitoelectric tidal field is stationary. We
confirm that in a calculation carried out to first order in the body's   
rotation, the exterior geometry bears no trace of this internal
dynamics; it remains stationary in spite of the time-dependent
interior.   
\end{abstract} 
\maketitle

\section{Introduction and summary} 
\label{sec:intro} 

A recent observation by Flanagan and Hinderer
\cite{flanagan-hinderer:08, hinderer:08}, that the tidal deformation
of neutron stars in inspiraling binaries could leave a measurable 
imprint on the gravitational waves they emit, has triggered a surge of   
activity to better understand the tidal deformation of compact bodies
in general relativity. This led to a precise formulation of
relativistic Love numbers \cite{damour-nagar:09,
  binnington-poisson:09, landry-poisson:14},  a computation of these
Love numbers for realistic models of neutron stars and a thorough
investigation of whether they can truly be measured in gravitational
waves \cite{hinderer-etal:10,  baiotti-etal:10, baiotti-etal:11,
  vines-flanagan-hinderer:11, pannarale-etal:11, lackey-etal:12,
  damour-nagar-villain:12,  read-etal:13, vines-flanagan:13,
  maselli-gualtieri-ferrari:13, lackey-etal:14, favata:14,
  yagi-yunes:14, delsate:15},  
the discovery of the remarkable $I$-Love-$Q$ relations 
\cite{yagi-yunes:13a, yagi-yunes:13b, doneva-etal:14,
  maselli-etal:13, yagi:14, haskell-etal:14, chakrabarti-etal:14}, 
and the computation of tidal invariants to be inserted within
point-particle actions to model the tidal response of an extended body 
\cite{bini-damour-faye:12,  chakrabarti-delsate-steinhoff:13a,
  chakrabarti-delsate-steinhoff:13b, dolan-etal:14, bini-damour:14}.  

To further contribute to this effort, we initiated a program to
calculate the tidal deformation of slowly rotating bodies in
general relativity. We began in Ref.~\cite{poisson:15} (Paper 0) with
the specific case of a black hole, and continued in
Ref.~\cite{landry-poisson:15a} (Paper I) with the determination of  
the exterior geometry of a material body; this objective was
simultaneously pursued by Pani {\it et al.}\ in
Ref.~\cite{pani-etal:15a}. In this paper we examine the interior
geometry of the material body, and determine the state of the fluid
making up the body. This objective was also pursued by Pani
{\it et al.}\ \cite{pani-etal:15b}, but these authors reach
dramatically different conclusions.     

Our main result is delightfully surprising. We find that {\it a slowly 
rotating body subjected to a stationary tidal field undergoes a
dynamical response, in which the fluid variables and the interior
metric vary with time}; the response occurs on the time scale of the
rotation period. We also confirm that {\it the exterior geometry
of the slowly rotating body bears no trace of this internal dynamics
 --- it is perfectly stationary}.  

These effects were not seen in Ref.~\cite{pani-etal:15b}. There are
two reasons for this discrepancy. The first and most important is that
these authors keep the fluid artificially static by demanding that the
velocity perturbation created by the tidal field vanishes. We require
instead the perturbation of the vorticity tensor to vanish, which is
accomplished by a nonzero velocity field. We argue below (and more
fully in Sec.~\ref{sec:fluid}) that our adopted state for the fluid is
the one naturally established when the body is assumed to begin in an 
unperturbed state. A second reason for the discrepancy is that these
authors consider tidal environments that are symmetric about the
body's rotation axis. As we shall see below (in Sec.~\ref{subsec:L2}),
the time dependence would be revealed in non-axisymmetric situations,
even under the static-fluid assumption. And indeed, consideration of
non-axisymmetric tidal fields reveals that the static state is
unphysical.\footnote{It is known that a non-axisymmetric tidal field
  exerts a torque that decreases the body's angular momentum over a
  very long time scale determined in part by the source of dissipation
  within the body. This is not the effect discussed here, which occurs
  over the short time scale of the rotation period. \label{foot:spindown}}    

To better explain the meaning of our results, we consider (for now)
the specific tidal environment created by a companion body of mass
$M'$ placed on a circular orbit of radius $b$, velocity $v$, and
angular velocity $\Omega_{\rm orb}$ around our reference body, which
has a mass $M$, radius $R$, and rotational angular velocity vector
$\Omega^a$. To leading order in the tidal interaction and in a
post-Newtonian expansion of the orbital motion, the tidal environment
is described by the quadrupole moments   
\begin{equation} 
\E_{ab} = -\frac{3 M'}{b^3} \Bigl( N_a N_b  - \frac{1}{3} \delta_{ab}
\Bigr), \qquad 
\B_{ab} = -\frac{3 M' v}{b^3} \Bigl( L_a N_b + N_a L_b \Bigr), 
\label{EB_circ} 
\end{equation} 
in which $N_a := [\cos(\Omega_{\rm orb} t), \sin(\Omega_{\rm orb} t),
0]$ points from the reference body to the companion, while 
$L_a := [0,0,1]$ is normal to the orbital plane. The tensor $\E_{ab}$
is known as the gravitoelectric tidal moment, and to leading order in
the tidal interaction, it provides a complete description of the tidal
environment in Newtonian gravity. The tensor $\B_{ab}$ is known as the
gravitomagnetic tidal moment, and it has no analogue in the Newtonian
theory. The phase of the tidal moments varies on the time scale of the
orbital period, and the amplitude varies on a much longer
radiation-reaction time scale. {\it In this work we assume that the
  orbital period is much longer than the body's rotational period.} In
this regime the body's response to the tidal field occurs over the
short time scale, and the variation of the tidal environment over the
long time scale can be neglected. This is the stationary regime
considered in this work. For our purposes we therefore take $\E_{ab}$
and $\B_{ab}$ to be time-independent, but we do not assume that they
take the specific forms displayed in Eq.~(\ref{EB_circ}); except for
the stationarity assumption, our tidal environment is generic. 

We create the tidal environment by immersing the body in a spacetime
that is not asymptotically flat, but modified to account for the
presence of $\E_{ab}$ and $\B_{ab}$. The details of the construction
are provided in Paper I \cite{landry-poisson:15a}, and our objective in 
this paper is to extend the exterior solution of Paper I to the body's
interior. This requires solving the Einstein field equations for the
metric, together with the relativistic Euler equation for the fluid
variables. This, in turn, requires the specification of the fluid's
state. For this purpose we rely on the fact that the amplitude of the
tidal moments $\E_{ab}$ and $\B_{ab}$ reflects an orbital evolution
over a very long radiation-reaction time scale. The reference body can
therefore be taken to be isolated in the remote past, and to have
begun in an unperturbed state in the absence of an initial tidal
field. This assumption is sufficient to specify the state of the fluid
at any later time: The Lagrangian perturbation 
$\Delta \omega_{\alpha\beta}$ of the vorticity tensor vanishes in the 
initial state, and conservation of vorticity guarantees that 
$\Delta \omega_{\alpha\beta}$ continues to vanish at later times. This
observation is a key physics input, and as we shall see in
Sec.~\ref{sec:fluid}, conservation of vorticity implies that fluid
motions develop as the tidal field gets established over the
radiation-reaction time scale.  

Solving the field and fluid equations reveals that a subset of metric
and fluid variables must depend on time, in spite of the fact that the
external tidal field is taken to be stationary. An example is 
\begin{equation} 
g_{tr} = t\, \kd_{tr1}(r)\, \Omega^a \B_{ab} n^b 
+ t\, \ko_{tr1}(r)\, \Omega_{\langle a} \B_{bc\rangle} n^a n^b n^c, 
\label{gtr} 
\end{equation} 
where $\kd_{tr1}(r)$ and $\ko_{tr1}(r)$ are functions of the radial
coordinate $r$ that will be determined below, $n^a := x^a/r$ is a unit
vector that points away from the body's center-of-mass, and the
angular brackets instruct to symmetrize all indices and remove all
traces. The first term in Eq.~(\ref{gtr}) is a dipole ($\ell=1$)
perturbation, while the second term is an octupole ($\ell=3$)
perturbation; these result from the coupling between the dipole, 
rotational perturbation generated by $\Omega^a$ and the quadrupole,
tidal perturbation generated by $\B_{ab}$. We find that all
time-dependent terms in the metric and fluid variables require the
existence of a gravitomagnetic field $\B_{ab}$. The gravitoelectric
field $\E_{ab}$ does not provoke a dynamical response; all associated
metric and fluid variables are time-independent. The time dependence
displayed in Eq.~(\ref{gtr}) describes a steady growth. We point out,
however, that our calculation is carried out to first order in $\Omega
:= |\Omega^a|$, and that it cannot distinguish between $\Omega t$ and
$\sin \Omega t$. We consider it likely that the time dependence is
actually bounded. 

The construction of Paper I guarantees that the exterior metric is
stationary. The time-dependent metric functions are therefore required 
to be supported inside the body only, and to vanish outside the
body. We show below (in Sec.~\ref{sec:FE}) that this is indeed
dictated by the fluid equations: the condition 
$\Delta p = 0$ at $r=R$, where $\Delta p$ is
the Lagrangian perturbation of the pressure, implies that $\kd_{tr1}(r)$
and $\ko_{tr1}(r)$ must smoothly vanish at $r=R$. We point out that
the stationary exterior is the consequence of a calculation carried
out to first order in $\Omega$; a calculation taken to higher order
would reveal the gravitational waves produced by the time-dependent
interior.    

The dynamical response of a slowly rotating body to a stationary tidal
field is a startling outcome. Before proceeding with a detailed
demonstration of this result, we attempt an intuitive explanation. The 
following argument is not meant to be rigorous or precise, but we
believe that it does capture the essence of the phenomenon.   

The dynamics of a perturbed fluid can be described in terms of a
Lagrangian displacement $\bm{\xi}$, which maps the position of a
perturbed fluid element to its original, unperturbed position. The
displacement field satisfies a complicated differential equation, and
this equation can be integrated by decomposing $\bm{\xi}$ in a
complete basis of normal modes $\bm{z}_\lambda$, with $\lambda$
denoting a mode label. Writing 
$\bm{\xi}(t,\bm{x}) = \sum_\lambda a_\lambda(t)
\bm{z}_\lambda(\bm{x})$ and inserting within the fluid equation
produces a mode equation of the form 
\begin{equation} 
\ddot{a}_\lambda + \omega_\lambda^2 a_\lambda = f_\lambda, 
\end{equation} 
where an overdot indicates differentiation with respect to $t$,
$\omega_\lambda$ is the mode's natural frequency, and 
$f_\lambda = \int \bm{F} \cdot \bm{z}_\lambda\, d^3x$ is the
forcing function, obtained by evaluating an overlap integral between  
the external perturbation $\bm{F}$ and each mode function. The
mode equation, of course, is the familiar equation that governs a
simple harmonic oscillator driven by an external force $f_\lambda$. 

In our context $f_\lambda$ is time-independent, and the expected 
stationary solution to the mode equation is obtained by balancing the
external force with the oscillator's restoring force; we get $a_\lambda
= \omega_\lambda^{-2} f_\lambda$. But what if the mode spectrum
includes zero-frequency modes? In this case there is no restoring
force to balance out the external force, and the solution can no
longer be stationary. Choosing an unperturbed state at $t=0$, we have
instead $a_\lambda = \frac{1}{2} f_\lambda\, t^2$. The displacement
field grows quadratically with time, and it gives rise to a velocity
field proportional to $t$. This is entirely analogous to our
situation; the velocity field found in Secs.~\ref{sec:fluid} and
\ref{sec:FE} does indeed grow linearly with time, just as $g_{tr}$ in
Eq.~(\ref{gtr}).   

Whence cometh the zero-frequency modes? The existence of such modes in 
relativistic fluids is well documented (see
Ref.~\cite{lockitch-andersson-friedman:00} for a clear presentation),
and we summarize the situation more fully in
Sec.~\ref{subsec:zero}. For the polar (even-parity) perturbations 
of a static, spherical body, there exists an infinite class of
zero-frequency $g$-modes describing a steady velocity field within the
body. For the axial (odd-parity) perturbations, there exists an
infinite class of zero-frequency $r$-modes, which also describe a
steady velocity field. We believe that these modes are implicated in
the tidal response of a slowly rotating body. In the absence of
rotation, the overlap integral between the external tidal field and
each zero-mode function vanishes, and the response is stationary. In
the presence of rotation, however, the forcing function no longer
vanishes, and the tidal response is dynamical. Our detailed
computations indicate that the overlap integral continues to vanish
when $\B_{ab} = 0$; the dynamical response requires a gravitomagnetic
tidal field.   

It seems plausible that the time-dependent metric and fluid variables
discussed in this work can be related to the rotational modes of
relativistic stars described by Lockitch, Andersson, and Friedman
\cite{lockitch-andersson-friedman:00,
  lockitch-friedman-andersson:03}; these generalize the famous
$r$-modes of Newtonian gravity. We shall not attempt a precise
identification, but the thought evokes a tantalizing
possibility. Like the $r$-modes \cite{andersson:98}, the rotational
modes are susceptible to the Chandrasekhar-Schutz-Friedman
instability \cite{chandrasekhar:70, friedman-schutz:78}, and this
suggests that an external tidal field can naturally create rotational
modes that will be driven unstable by the emission of gravitational
radiation. This could be a significant factor in the tidal interaction
of neutron stars. We shall not, however, speculate any further.     

Our main objective with this paper is to demonstrate the existence of
the dynamical response of a slowly rotating body to a stationary tidal
field. Because the dynamical response is associated exclusively with
the gravitomagnetic field $\B_{ab}$, we shall ignore the
gravitoelectric field $\E_{ab}$ altogether. We begin our technical 
developments in Sec.~\ref{sec:spacetime}, where we introduce the
background metric of a static, spherically symmetric body, set up the  
rotational and tidal perturbations, and describe the coupling
between these perturbations. In Sec.~\ref{sec:fluid} we introduce the
fluid variables, and work out the consequences of the key fluid
equation, $\Delta \omega_{\alpha\beta} = 0$. In Sec.~\ref{sec:FE} we
manipulate the Einstein field equations to deliver a complete set of
equations for all metric and fluid variables. We leave the task of
actually integrating these equations to a forthcoming
publication.\footnote{P.\ Landry and E.\ Poisson, in preparation.}   

\section{Spacetime of a tidally deformed, slowly rotating body} 
\label{sec:spacetime} 

The metric of a tidally deformed, slowly rotating body is
constructed as a perturbation of the background metric of an
undeformed and nonrotating body. It is assumed that the unperturbed
body is in hydrostatic equilibrium, and the background metric is
expressed in the standard form  
\begin{equation} 
ds^2 = -e^{2\psi}\, dt^2 + f^{-1}\, dr^2 + r^2\, d\Omega^2, 
\label{background} 
\end{equation} 
where $\psi = \psi(r)$, $f = 1-2m(r)/r$, and $d\Omega^2 := d\theta^2 +
\sin^2\theta\, d\phi^2$. The Einstein field equations imply that
$\psi$ and $m$ are solutions to 
\begin{equation} 
\frac{dm}{dr} = 4\pi r^2 \mu, \qquad 
\frac{d \psi}{dr} =\frac{m + 4\pi r^3 p}{r^2 f}, 
\end{equation} 
where $\mu$ is the fluid's energy density (the sum of mass density
$\rho$ and density of internal energy $\epsilon$), and $p$ is the
pressure. The condition of hydrostatic equilibrium is 
\begin{equation} 
\frac{dp}{dr} = - \frac{(\mu+p)(m + 4\pi r^3 p)}{r^2 f}. 
\end{equation} 
These equations must be supplemented with an equation of state, which
we take to be of the form $p = p(\rho)$ and 
$\epsilon = \epsilon(\rho)$; the fluid is thus assumed to have a
constant specific entropy.  The interior metric must be matched to the
Schwarzschild exterior metric at $r=R$, at which $p = 0$.  

The slow rotation of the body is incorporated as a linear perturbation
of the background metric. We assume that the rotation is rigid, and
the metric perturbation $p_{\alpha\beta}$ has the single nonvanishing
component\footnote{Our function $\Omega \omega(r)$ is typically
  denoted $\bar{\omega}(r)$ in the literature.}  
\begin{equation} 
p^{\rm rotation}_{t\phi} = -\Omega (1-\omega) r^2 \sin^2\theta, 
\label{rotation} 
\end{equation} 
where $\Omega$ is the body's angular velocity, and $\omega(r)$ is a
solution to 
\begin{equation} 
r^2 f \frac{d^2\omega}{dr^2} 
+ \bigl[ 4f - 4\pi r^2 (\mu+p) \bigr] r \frac{d\omega}{dr}  
- 16\pi r^2 (\mu+p) \omega = 0. 
\label{omega} 
\end{equation} 
The interior solution is matched at $r=R$ to the exterior solution  
$\omega_{\rm ext} = 1 - 2I/r^3$, where $I := J/\Omega$ is the body's 
moment of inertia. 

As was explained in Sec.~\ref{sec:intro}, we place the body in a tidal
environment characterized by a gravitomagnetic quadrupole   
moment $\B_{ab}$, a Cartesian symmetric-tracefree (STF) tensor that
we take to be independent of time --- the tidal environment is
stationary. The five independent components of $\B_{ab}$ can be
packaged in spherical-harmonic coefficients $\Bq_\m$, with the label
$\m$ ranging over the five values associated with spherical harmonics
of degree $\ell=2$. The precise packaging is described in Table I of
Paper I \cite{landry-poisson:15a}.  

The tidal perturbation is conveniently decomposed in spherical
harmonics. Because the perturbation created by $\B_{ab}$ is axial
(odd-parity) in nature, the decomposition is accomplished with the
odd-parity vectorial harmonics    
\begin{equation} 
X_A^{\ell\m} := -\epsilon_A^{\ B} D_B Y^{\ell\m}.  
\label{odd_harm} 
\end{equation}  
Here, uppercase Latin indices $A, B, C, \cdots$ range over the
angular variables $\theta$ and $\phi$, $Y^{\ell\m}$ are the usual 
spherical-harmonic functions (with the normalization adopted in 
Table II of Paper I), $\Omega_{AB} = \mbox{diag}(1,\sin^2\theta)$ is
the metric on a unit two-sphere, $D_A$ is the covariant-derivative
operator compatible with this metric, and $\epsilon_{AB}$ is the
Levi-Civita tensor on the unit two-sphere 
($\epsilon_{\theta\phi} = \sin\theta$); it is our convention that an 
uppercase Latin index is raised with $\Omega^{AB}$, the matrix inverse
to $\Omega_{AB}$. 

We adopt the Regge-Wheeler gauge, and find (see
Ref.~\cite{landry-poisson:15b} for a derivation) that the only
nonvanishing component of the tidal perturbation is given 
by\footnote{Our notation for $\bq_t(r)$ differs from the one
  adopted in Sec.~IV of Paper I; we have that 
  $\bq_t[\text{here}] = \frac{2}{3} r^3 \bq_t[\text{Paper I}]$.}   
\begin{equation} 
p^{\rm tidal}_{tA} = \bq_t(r) \Bq_A,
\label{tidal}
\end{equation} 
where
\begin{equation} 
\Bq_A := \frac{1}{2} \sum_\m \Bq_\m X_A^{2\m}.  
\end{equation} 
The Einstein field equations imply that the radial function $\bq_t$
satisfies 
\begin{equation} 
r^2 f \frac{d^2 \bq_t}{dr^2} 
- 4\pi r^3 (\mu+p) \frac{d\bq_t}{dr}  
- 2\bigl[ 3-2m/r \mp 4\pi r^2(\mu+p) \bigr] \bq_t = 0. 
\label{bq_eqn} 
\end{equation} 
The choice of sign in this equation depends on the assumed state 
of the fluid; the upper sign corresponds to an irrotational fluid with
internal motions, while the lower sign corresponds to a static fluid
with no internal motions. As we shall explain below, the upper sign is
the appropriate choice for our purposes in this work.  

We next allow the (dipole) rotational perturbation of
Eq.~(\ref{rotation}) and the (quadrupole) tidal perturbation of
Eq.~(\ref{tidal}) to source a second-order perturbation. We continue
to work to first order in $\Omega$ and to first order in $\B_{ab}$,
but we introduce terms of order $\Omega \B_{ab}$ in the perturbed
metric. The composition of the $\ell = 1$ and $\ell = 2$ spherical
harmonics is reflected in the bilinear moments\footnote{We use the
  same notation as in Paper I for the bilinear moments, but construct
  them with $\Omega^a$ instead of $\chi^a$, the dimensionless
  angular-momentum vector. The moments therefore differ by a factor
  $\chi/\Omega = I/M^2$. \label{foot:4}}  
\begin{equation} 
\K_a := \B_{ab} \Omega^b, \qquad  
\hat{\B}_{ab} := 2 \Omega^c \epsilon_{cd(a} \B^d_{\ b)}, \qquad 
\K_{abc} := \B_{\langle a b} \Omega_{c\rangle}, 
\end{equation} 
where $\Omega_a := [0,0,\Omega]$ is the angular-velocity vector,
$\epsilon_{abc}$ is the antisymmetric permutation symbol, the angular
brackets indicate the STF operation (symmetrize all indices and remove
all traces), and all lowercase Latin indices are raised and lowered
with the Euclidean metric. The independent components of the STF
tensors $\K_a$, $\hat{\B}_{ab}$, and $\K_{abc}$ can also be packaged
in spherical-harmonic coefficients $\Kd_\m$, $\Bhatq_\m$, and
$\Ko_\m$, respectively; the precise definitions are displayed in 
Table I of Paper I (with the change of notation described in footnote
\ref{foot:4}).  

We continue to adopt the Regge-Wheeler gauge for the second-order
perturbation, and for the $\ell = 1$ terms generated by $\K_a$ we
adopt the extension formulated by Campolattaro and Thorne
\cite{campolattaro-thorne:70}. We observe that $\K_a$ and $\K_{abc}$
create polar (even-parity) perturbations, while $\hat{\B}_{ab}$
creates axial (odd-parity) perturbations, and express the perturbation
as\footnote{In addition to the change of notation described previously
  for the bilinear moments, we note that our notation for the various
  functions differs from the one adopted in Paper I. We have 
  $\kd_{tt}[\text{here}] = r^2 \kd_{tt}[\text{Paper I}]$,
  $\kd_{rr}[\text{here}] = r^2 \kd_{rr}[\text{Paper I}]$,
  $\ko_{tt}[\text{here}] = -r^2 \ko_{tt}[\text{Paper I}]$,
  $\ko_{rr}[\text{here}] = -r^2 \ko_{rr}[\text{Paper I}]$,
  $\ko[\text{here}] = r^4 \ko[\text{Paper I}]$, 
  $\bhatq_t[\text{here}] = -r^3 \bhatq_t[\text{Paper I}]$, and 
  $\bhatq_r[\text{here}] = -r^3 \bhatq_r[\text{Paper I}]$. \label{foot:5}}  
\begin{subequations} 
\label{coupled} 
\begin{align} 
p^{\rm bilinear}_{tt} &= \kd_{tt}(t,r) \Kd + \ko_{tt}(t,r) \Ko, \\ 
p^{\rm bilinear}_{tr} &= \kd_{tr}(t,r) \Kd + \ko_{tr}(t,r) \Ko, \\ 
p^{\rm bilinear}_{rr} &= \kd_{rr}(t,r) \Kd + \ko_{rr}(t,r) \Ko, \\ 
p^{\rm bilinear}_{tA} &= \bhatq_t(t,r) \Bhatq_A, \\ 
p^{\rm bilinear}_{rA} &= \bhatq_r(t,r) \Bhatq_A, \\ 
p^{\rm bilinear}_{AB} &= \ko(t,r) \Omega_{AB} \Ko, 
\end{align} 
\end{subequations} 
where 
\begin{equation} 
\Kd = \sum_\m \Kd_\m Y^{1\m}, \qquad 
\Ko = \sum_\m \Ko_\m Y^{3\m}, \qquad 
\Bhatq_A = \frac{1}{2} \sum_\m \Bhatq_\m X^{2\m}_A 
= -\Omega \partial_\phi \Bq_A.  
\end{equation} 
It should be noticed that we allow $\{ \kd_{tt}, \cdots, \ko \}$ to be
functions of both $t$ and $r$, and that $\Bhatq_A$ vanishes when the
tidal environment is axisymmetric.   

\section{Perturbed fluid} 
\label{sec:fluid} 

The perturbations created by $\B_{ab}$, $\K_a$, $\hat{B}_{ab}$, and 
$\K_{abc}$ disturb the fluid, and we give this disturbance a
Lagrangian formulation as summarized, for example, in Sec.~2.2 of
Ref.~\cite{friedman-stergioulas:13}. We assume that the fluid is
barotropic, which means that the equation of state of the perturbed
fluid is the same as for the unperturbed fluid.  

The Lagrangian displacement vector $\xi_\alpha$ is decomposed as  
\begin{subequations} 
\begin{align} 
\xi_r &= \xid_r(t,r) \Kd + \xio_r(t,r) \Ko, \\
\xi_A &= \xiq(t,r) \Bq_A + \xid(t,r) \Kd_A 
+ \xihatq(t,r) \Bhatq_A + \xio(t,r) \Ko_A, 
\end{align} 
\end{subequations} 
and the Eulerian perturbation of the velocity vector is decomposed as 
\begin{subequations} 
\begin{align} 
\delta u_r &= \vd_r(t,r) \Kd + \vo_r(t,r) \Ko, \\ 
\delta u_A &= \vq(r) \Bq_A + \vd(t,r) \Kd_A 
+ \vhatq(t,r) \Bhatq_A + \vo(t,r) \Ko_A; 
\end{align} 
\end{subequations} 
the time component of $\xi_\alpha$ plays no role in our discussion,
and $\delta u_t$ can be related to the other components by properly
normalizing the perturbed velocity vector. The various expansion
coefficients are related by  
\begin{subequations} 
\begin{align} 
\vq &= e^{-\psi} \bigl( \partial_t \xiq + \bq_t \bigr),  
\label{vq_vs_xi} \\ 
\vd_r &= e^{-\psi} \bigl( \partial_t \xid_r + \kd_{tr} \bigr), \\ 
\vd &= e^{-\psi} \partial_t \xi^d, \\ 
\vhatq &= e^{-\psi} \bigl( \partial_t \xihatq - \xiq + \bhatq_t \bigr), \\  
\vo_r &= e^{-\psi} \bigl( \partial_t \xio_r + \ko_{tr} \bigr), \\ 
\vo &= e^{-\psi} \partial_t \xio. 
\end{align} 
\end{subequations} 
The Eulerian perturbation of the pressure is expressed as  
\begin{equation} 
\delta p = \pd(t,r) \Kd + \po(t,r) \Ko, 
\end{equation} 
and the barotropic assumption ensures that the perturbations in energy
density $\mu$ and specific enthalpy $h$ are given by  
$\delta \mu = (d\mu/dp) \delta p$ and 
$\delta h = h(\mu+p)^{-1} \delta p$, respectively. We recall that the
Lagrangian perturbation $\Delta Q$ of a fluid variable $Q$ is related
to its Eulerian perturbation $\delta Q$ by $\Delta Q = \delta Q 
+ {\cal L}_\xi Q$, where ${\cal L}_\xi$ indicates Lie differentiation
in the direction of $\xi^\alpha$.  

The state of the fluid is constrained by the fact that the vorticity
tensor $\omega_{\alpha\beta} := \nabla_\alpha (h u_\beta) -
\nabla_\beta (h u_\alpha)$ is conserved along the fluid world lines:  
${\cal L}_u \omega_{\alpha\beta} = 0$; see, for example, Sec.~1.1 of
Ref.~\cite{friedman-stergioulas:13} for a derivation. Taking a
Lagrangian perturbation of this equation gives rise to  
${\cal L}_u \Delta \omega_{\alpha\beta} = 0$, 
the statement that $\Delta \omega_{\alpha\beta}$ is conserved along
the fluid world lines. To implement this condition we imagine that in
spite of our assumption of stationarity, the tidal
perturbation was switched on adiabatically in the remote past, so that
the fluid began in an unperturbed state.\footnote{For our purposes in this
discussion, the body's rotation is excluded from the perturbation and
included in the unperturbed configuration; the perturbation refers
exclusively to the tidal field.} In this initial state 
$\Delta \omega_{\alpha\beta} = 0$, and the conservation statement 
implies that $\Delta \omega_{\alpha\beta}$ continues to vanish on each 
world line. Conservation of vorticity therefore guarantees that 
\begin{equation} 
\Delta \omega_{\alpha\beta} = 0 
\label{novort} 
\end{equation} 
at any time throughout the fluid. 

The conditions of Eq.~(\ref{novort}) have far-reaching
consequences. We note first that the angular components of this
equation imply that 
\begin{equation} 
\vq = 0. 
\label{vq} 
\end{equation} 
This, we recall, is the piece of $\delta u_A$ associated with the
tidal perturbation of a nonrotating body, and the vorticity constraint
implies the existence of internal motions within the fluid
\cite{favata:06, landry-poisson:15b}. These motions are described by   
$\delta u^A = -\frac{1}{2} r^{-2} e^{-\psi} \bq_t(r) \Omega^{AB}
\Bq_B$, and they are gradually established as the tidal field is
adiabatically switched on. The assignment of Eq.~(\ref{vq}) dictates 
the choice of upper sign in Eq.~(\ref{bq_eqn}). From $\vq = 0$ and
Eq.~(\ref{vq_vs_xi}) we next find that $\xiq = -t\, \bq_t(r)$.   

The angular components of Eq.~(\ref{novort}) also produce $\vhatq =
-\frac{1}{3} e^{-\psi} \omega\, \xiq$, in which we insert our previous
result for $\xiq$. We arrive at 
\begin{equation} 
\vhatq = \frac{1}{3} t\, e^{-\psi} \omega(r) \bq_t(r), 
\label{vhatq} 
\end{equation} 
the striking statement that {\it the response of a slowly rotating
  body to a stationary tidal field is necessarily dynamical}. The $rA$
components of 
Eq.~(\ref{novort}) give 
\begin{subequations} 
\label{vd_vo} 
\begin{align} 
\vd_r &= \frac{3}{5} t\, e^{-\psi} \biggl[ \omega \frac{d\bq_t}{dr} 
+ 2 \biggl( \frac{d\omega}{dr} 
+ \frac{r - 4m - 8\pi r^3 p}{r^2 f} \omega \biggr) \bq_t \biggr]  
+ \partial_r \vd - \frac{m + 4\pi r^3 p}{r^2 f} \vd, \\ 
\vo_r &= \frac{1}{3} t\, e^{-\psi} \biggl[ 2\omega \frac{d\bq_t}{dr} 
- \biggl( \frac{d\omega}{dr} 
+ \frac{6r - 14m - 8\pi r^3 p}{r^2 f} \omega \biggr) \bq_t \biggr]  
+ \frac{1}{3} \partial_r \vo 
- \frac{m + 4\pi r^3 p}{3 r^2 f} \vo, 
\end{align} 
\end{subequations} 
and these also reveal the time dependence of the velocity field. The
$tA$ components of Eq.~(\ref{novort}) relate the pressure
perturbations to other variables; we have 
\begin{subequations} 
\label{pd_pd} 
\begin{align} 
\pd &= e^{-\psi} (\mu+p) \biggl[ \frac{1}{2} e^{-\psi} \kd_{tt} 
- \frac{3}{5} e^{-\psi} (1+\omega) \bq_t 
- \partial_t \vd \biggr], \\ 
\po &= e^{-\psi} (\mu+p) \biggl[ \frac{1}{2} e^{-\psi} \ko_{tt} 
+ \frac{1}{3} e^{-\psi} (3-2\omega) \bq_t 
- \frac{1}{3} \partial_t \vo \biggr]. 
\end{align} 
\end{subequations} 
This exhausts the information disclosed by Eq.~(\ref{novort}); the 
remaining $tr$ component is redundant.  

\section{Field equations} 
\label{sec:FE} 

\subsection{Zero-frequency modes}  
\label{subsec:zero} 

Before we proceed with an analysis of the Einstein field equations, we
briefly recall the existence of two classes of zero-frequency modes
for the perturbations of a static, spherically symmetric body. For the 
purposes of this discussion we set $\Omega = 0$, switch off the
external tidal field, and turn off the time-dependence of the
perturbation variables. We summarize the presentation contained in 
Sec.~III of Ref.~\cite{lockitch-andersson-friedman:00}. 

We first consider a polar (even-parity) perturbation described by the
metric variables $\{ k_{tt}, k_{tr}, k_{rr}, k \}$ and the fluid
variables $\{ v_r, v, p \}$; we suppress the multipole label (such as
${\sf d}$ and ${\sf o}$) to emphasize the fact that the discussion is
not limited to the dipole and octupole perturbations examined in this
work. The variables decouple into the groups  
$\{ k_{tt}, k_{rr}, k, p \}$ and $\{ k_{tr}, v_r, v \}$, and it can be
shown that for a homogeneous perturbation (not driven by an external
tidal field), the variables belonging to the first group vanish. When
the fluid is barotropic, however, the variables belonging to the
second group admit an infinity of solutions, each one characterized by
a freely specifiable $v_r$. These define the class of zero-frequency
$g$-modes, which were first identified by Thorne \cite{thorne:69}.  

We next consider an axial (odd-parity) perturbation described by the
metric and fluid variables $\{\hat{b}_t, \hat{b}_r, \hat{v} \}$,
where we again suppress the multipole label (such as ${\sf q}$). In
this case the field equations reveal the existence of another infinity
of solutions, each one characterized by a freely specifiable $\hat{v}$
and a vanishing $\hat{b}_r$. These define the class of zero-frequency
$r$-modes, which are not limited to barotropic fluids. 

\subsection{Field equations: $\ell = 2$} 
\label{subsec:L2} 

We substitute the metric of Sec.~\ref{sec:spacetime} and the fluid 
variables of Sec.~\ref{sec:fluid} into the Einstein field equations 
\begin{equation} 
G_{\alpha\beta} = 8\pi T_{\alpha\beta},  
\end{equation} 
expand to first order in $\Omega$ and $\Bq_\m$, and decompose each
component in (scalar, vector, and tensor) spherical harmonics. The
field equations for each multipole order decouple, and in this
subsection we examine the $\ell=2$ sector associated with the
variables $\{ \bhatq_t, \bhatq_r, \vhatq \}$. 

The velocity perturbation is eliminated with Eq.~(\ref{vhatq}), and
inspired by this relation, we assume that the metric variables can be
at most linear in $t$, so that $\bhatq_t = \bhatq_{t0}(r) + t\, 
\bhatq_{t1}(r)$ and $\bhatq_r = \bhatq_{r0}(r) + t\, 
\bhatq_{r1}(r)$. The $rA$ components of the field equations then imply
that $\bhatq_{r1} = 0$, and the $tA$ components give rise to a
homogeneous differential equation for $\bhatq_{t0}$. The solution to
this equation represents an $r$-mode, and to simplify our solution we
choose to eliminate this degree of freedom by setting 
$\bhatq_{t0} = 0$; the $r$-mode can be restored at will.      

We are left with 
\begin{equation} 
\bhatq_t(t,r) = t\, \bhatq_{t1}(r), \qquad 
\bhatq_r(t,r) = \bhatq_{r0}(r), 
\end{equation} 
and the $tA$ components of the field equations give rise to 
\begin{equation} 
r^2 f \frac{d^2 \bhatq_{t1}}{dr^2} 
- 4\pi r^3 (\mu+p)  \frac{d \bhatq_{t1}}{dr}
- \frac{2}{r} \bigl[ 3r - 2m - 4\pi r^3(\mu+p) \bigr] \bhatq_{t1} 
- \frac{16\pi}{3} r^2 (\mu+p) \omega \bq_t = 0,  
\label{bhatq_t} 
\end{equation} 
a differential equation that determines $\bhatq_{t1}(r)$. The $rA$
components return 
\begin{equation} 
\bhatq_{r0} = \frac{1}{4} e^{-2\psi}  \biggl[ 
r^2 \frac{d \bhatq_{t1}}{dr}  - 2 r \bhatq_{t1} 
- r^2 (1-\omega) \frac{d\bq_t}{dr} 
- \frac{1}{3} r \biggl( r \frac{d\omega}{dr} 
+ 6\omega - 6 \biggr) \bq_t \biggr],  
\label{bhatq_r} 
\end{equation} 
an algebraic equation for $\bhatq_{r0}(r)$. 

The interior metric variables must be matched with the exterior
solutions at $r=R$. The exterior metric was constructed in Paper I
\cite{landry-poisson:15a}, and we have re-examined this construction
to ensure that we didn't incorrectly eliminate time-dependent
terms. The conclusion of this exercise is that the metric of Paper I
requires no change; {\it the exterior metric must be stationary, in
  spite of the  time-dependent interior}. This implies that
$\bhatq_{t1}$ must satisfy the boundary conditions\footnote{Equation
  (\ref{surface}) may seem to provide too many conditions. This is not
  so. We assume that the equation of state approaches a polytropic
  form near the surface, so that  $p \propto \mu^{1+1/n}$, with 
  $n > 0$. The structure equations then imply that as $r \to R$, $p$
  and $\mu$ approach zero as $p \propto (1-r/R)^{n+1}$ and 
  $\mu \propto (1-r/R)^n$. Inspection of Eq.~(\ref{bhatq_t}) finally
  reveals that if $\bhatq_{t1}$ is to vanish at $r=R$, it must do so
  as $(1-r/R)^{2+n}$. \label{foot:surface}}    
\begin{equation} 
\bhatq_{t1} = 0 = \frac{d \bhatq_{t1}}{dr}  \qquad 
\text{at $r=R$}, 
\label{surface} 
\end{equation} 
and that $\bhatq_{r0}$ must match at $r = R$ the value listed in Table
IV of Paper I (after taking into account the change of notation
documented in footnotes \ref{foot:4} and \ref{foot:5}). In addition,
these functions must satisfy regularity conditions at
$r=0$.

It is interesting to reconsider the interior solution when the fluid
is taken to be static, in the sense that it is prevented to move, in
contradiction with our assumption in Eq.~(\ref{novort}). In this case
we must set $\vq = e^{-\psi} b_t$ and $\vhatq = e^{-\psi} \bhatq_t$ to
ensure that $\delta u^A = 0$, and the conservation equations  
$\nabla_\beta T^{\alpha\beta} = 0$ imply that $\bhatq_t$ must be
of the form $\bhatq_t(t,r) = \bhatq_{t0}(r) + t\, \bq_t(r)$; it is
significant that the time-dependent term involves the driving function
$\bq_t(r)$. The field equations further imply that $\bhatq_{t0}(r)$
satisfies a homogeneous differential equation; the solution represents
another $r$-mode, which we again have the freedom to eliminate. The
matching with the exterior solution creates a contradiction: The
interior expression for $\bhatq_t$ is necessarily time-dependent and
does not vanish at $r=R$, but the exterior expression is necessarily
stationary. We conclude that {\it a static fluid is an unphysical
  configuration when a slowly rotating body is deformed by a
  stationary tidal field}. We note that the contradiction is lifted
when the tidal field is axisymmetric; in this case $\Bhatq_A = 0$ and
the variables $\{ \bhatq_t, \bhatq_r, \vhatq \}$ are not defined. It
is for this reason that the contradiction was not noticed by the
authors of Ref.~\cite{pani-etal:15b}, who restricted their attention
to axisymmetric tidal fields.\footnote{Recall footnote
  \ref{foot:spindown}. The time dependence discussed here occurs over
  the short time scale of the rotation period. It has nothing to do
  with the tidal torquing of the body, which occurs over a much longer
  time scale.}  

\subsection{Field equations: $\ell = 1$} 
\label{subsec:L1} 

In this subsection we examine the $\ell=1$ sector of the perturbation,
associated with the variables $\{ \kd_{tt}, \kd_{rr}, \pd \}$ and 
$\{ \kd_{tr}, \vd_r, \vd \}$. The variables $\vd_r$ and $\pd$ are
eliminated with Eqs.~(\ref{vd_vo}) and (\ref{pd_pd}), respectively,
and we find that the $tA$ and $rA$ components of the field equations
allow us to further eliminate $\vd$ and $\kd_{rr}$. We are left with
$\kd_{tt}$ from the first group of variables, and $\kd_{tr}$ from the
second group. We next assume the explicit forms 
$\kd_{tt}(t,r) = \kd_{tt0}(r)$ and 
$\kd_{tr}(t,r) = \kd_{tr0}(r) + t\, \kd_{tr1}(r)$, which ensure that
all variables from the first group are time-independent, while those
from the second group are linear in time. Then the $tr$ component of
the field equations implies that $\kd_{tr0}$ satisfies a homogeneous
differential equation. The solution to this equation represents a
$g$-mode, and we simplify our solution by setting $\kd_{tr0} = 0$; the
freedom to incorporate the $g$-mode can be restored at will.    

We are left with 
\begin{equation}  
\kd_{tt}(t,r) = \kd_{tt0}(r), \qquad 
\kd_{tr}(t,r) = t\, \kd_{tr1}(r), 
\end{equation} 
and these functions are determined by the $tr$ and $rr$ components of
the field equations, respectively.  We have 
\begin{align} 
0 &= r^2 f \frac{d^2 \kd_{tr1}}{dr^2} 
+ \biggl[ 3 (m-4\pi r^3 \mu) + (m + 4\pi r^3 p) \frac{d\mu}{dp}
\biggr] \frac{d \kd_{tr1}}{dr}  
\nonumber \\ & \quad \mbox{} 
- \frac{2}{r^2 f} \biggl\{ 
\bigl[ 1 - 10 \pi r^2 (\mu + p) + 16\pi^2 r^4 p^2 \bigr] r^2   
+ 4\pi r^3 (5\mu+7p) m - 3m^2 
- (m+4\pi r^3 p)^2 \frac{d\mu}{dp} \biggr\} \kd_{tr1} 
\nonumber \\ & \quad \mbox{} 
- \frac{48\pi}{5} r^2(\mu+p) \omega \frac{d \bq_t}{dr} 
- \frac{96\pi}{5}  (\mu+p) \biggl[ r^2 \frac{d\omega}{dr} 
+ \frac{r - 4m - 8\pi r^3 p}{f} \omega \biggr] b_t    
\end{align} 
and 
\begin{align} 
0 &=  r \frac{d \kd_{tt0}}{dr} 
+ \frac{2 \bigl[ m - 2\pi r^3(\mu+p) \bigr](r - m + 4\pi r^3 p)}
  {r f (m+4\pi r^3 p)} \kd_{tt0} 
- \frac{ r^2(r-m+4\pi r^3 p)}{2(m+4\pi r^3 p)} \frac{d \kd_{tr1}}{dr} 
\nonumber \\ & \quad \mbox{} 
+ \frac{ \bigl[ 1 + 2\pi r^2(1 + 4\pi r^2 p)(\mu-p) \bigr] r^2
- \bigl[ 5 + 2\pi r^2 (\mu+p) \bigr] r m + 5m^2}
{f (m+ 4\pi r^3 p)} \kd_{tr1} 
\nonumber \\ & \quad \mbox{} 
- \frac{3r}{10} \biggl[ 
\frac{r(r-m+4\pi r^3 p)}{m+4\pi r^3 p} \frac{d\omega}{dr} 
- 4 \omega + 4 \biggr] \frac{d \bq_t}{dr} 
\nonumber \\ & \quad \mbox{} 
- \frac{3(r-m+4\pi r^3 p)}{5(m+4\pi r^3 p)}  \biggl[ 
r \frac{d\omega}{dr} 
- 4 \frac{m + 2\pi r^3(\mu+p)}{r f} \omega 
+ 4 \frac{m - 2\pi r^3(\mu+p)}{r f} \biggr] \bq_t. 
\end{align}   
The remaining variables are given by 
\begin{subequations} 
\begin{align} 
\kd_{rr} &= e^{-2\psi} \biggl[ 
- \frac{r^2}{r-m+4\pi r^3 p} \frac{d \kd_{tt0}}{dr} 
+ \frac{1}{f} \kd_{tt0} 
+ \frac{r^2}{r-m+4\pi r^3 p} \kd_{tr1} 
+ \frac{6(1-\omega) r^2}{5(r-m+4\pi r^3 p)} \frac{d\bq_t}{dr} 
- \frac{6(1-\omega)}{5f} \bq_t \biggr], \\ 
\pd &= e^{-2\psi} \biggl[ 
\frac{f}{16\pi} \frac{d \kd_{tr1}}{dr}  
+ \frac{m-2\pi r^3(\mu-p)}{8\pi r^2} \kd_{tr1} 
+ \frac{1}{2} (\mu+p) \kd_{tt0} 
- \frac{3}{5} (\mu+p)(1+\omega) \bq_t \biggr],  
\end{align}
\end{subequations}  
and 
\begin{subequations} 
\begin{align} 
\vd_r &= -t\, \frac{ e^{-\psi} \bigl[ 1 - 8\pi r^2(\mu+p) \bigr] }  
{ 8\pi r^2 (\mu+p) } \kd_{tr1}, \\  
\vd &= -t\, \frac{ e^{-\psi} }{16\pi r^2 (\mu+p)} \biggl\{ 
r^2 f \frac{d \kd_{tr1}}{dr} 
+ 2 \bigl[ m - 2\pi r^3(\mu-p) \bigr] \kd_{tr1} \biggr\}. 
\end{align} 
\end{subequations} 

The interior solutions must be matched with the exterior solutions of
Paper I at $r=R$, and this reveals that $\kd_{tr1}$ must satisfy the
boundary conditions\footnote{The calculation carried out in footnote
  \ref{foot:surface} also reveals that $\kd_{tr1} \propto
  (1-r/R)^{2+n}$ when $r \to R$.}  
\begin{equation} 
\kd_{tr1} = 0 = \frac{d \kd_{tr1}}{dr} \qquad 
\text{at $r=R$}.  
\end{equation} 
These conditions also follow from the requirement that $\Delta p = 0$
at $r= R$, assuming that both $\mu$ and $p$ go to zero at the surface. 
The matching also requires $\kd_{tt0}$ and $\kd_{rr}$ to agree at
$r=R$ with the values listed in IV of Paper I (after taking into
account the change of notation documented in footnotes \ref{foot:4}
and \ref{foot:5}).  In addition, all these functions must satisfy
regularity conditions at $r=0$.

We again reconsider the interior solution to allow the fluid to be
static. We now set $\vd_r = e^{-\psi} \kd_{tr}$ and $\vd = 0$ to  
ensure that $\delta u^r = 0 = \delta u^A$. The conservation  
equations then determine $\pd$ and imply that $\kd_{tr}$ must be
time-independent. We next assume that the remaining metric variables
are stationary, and verify that the field equations are consistent
with this assumption. Our conclusion, that it is possible to
construct a stationary solution to the $\ell = 1$ interior problem
when the fluid is static, is compatible with the computations 
presented in Ref.~\cite{pani-etal:15b}.  

\subsection{Field equations: $\ell = 3$} 
\label{subsec:L3} 

Finally we examine the $\ell=3$ sector of the perturbation, associated
with the variables $\{ \ko_{tt}, \ko_{rr}, \ko, \po \}$ and 
$\{ \ko_{tr}, \vo_r, \vo \}$. The variables $\vo_r$ and $\po$ are
eliminated with Eqs.~(\ref{vd_vo}) and (\ref{pd_pd}), respectively,
and we find that the $tA$ and $AB$ components of the field equations
allow us to further eliminate $\vo$ and $\ko_{rr}$. We are left with
$\ko_{tt}$ and $\ko$ from the first group of variables, and $\ko_{tr}$
from the second group. We then assume the explicit forms 
$\ko_{tt}(t,r) = \ko_{tt0}(r)$, $\ko(t,r) = \ko_0(r)$, and 
$\ko_{tr}(t,r) = \ko_{tr0}(r) + t\, \ko_{tr1}(r)$. The $tr$ component
of the field equations implies that $\ko_{tr0}$ satisfies a
homogeneous differential equation, and its solution represents a
$g$-mode that we allow ourselves to discard.  

We are left with 
\begin{equation}  
\ko_{tt}(t,r) = \ko_{tt0}(r), \qquad \ko(t,r) = \ko_0(r), \qquad
\ko_{tr}(t,r) = t\, \ko_{tr1}(r), 
\end{equation} 
and the differential equations satisfied by these functions are
obtained from the $tr$, $rr$, and $rA$ components of the field
equations, respectively. We find 
\begin{align} 
0 &= r^2 f \frac{d^2 \ko_{tr1}}{dr^2} 
+ \biggl[ 3 (m-4\pi r^3 \mu) + (m + 4\pi r^3 p) \frac{d\mu}{dp}
\biggr] \frac{d \ko_{tr1}}{dr}  
\nonumber \\ & \quad \mbox{} 
- \frac{2}{r^2 f} \biggl\{ 
2 \bigl[ 3 - 5\pi r^2 (\mu+p) + 8\pi^2 r^4 p^2 \bigr] r^2 
- 2 \bigl[ 5 - 2\pi r^2 (5\mu + 7p) \bigr] r m - 3 m^2    
- (m+4\pi r^3 p)^2 \frac{d\mu}{dp} \biggr\} \ko_{tr1} 
\nonumber \\ & \quad \mbox{} 
- \frac{32\pi}{3} r^2(\mu+p) \omega \frac{d \bq_t}{dr} 
+ \frac{16\pi}{3} (\mu+p) \biggl[ r^2 \frac{d\omega}{dr} 
+ 2\frac{3r - 7m - 4\pi r^3 p}{f} \omega \biggr] b_t     
\end{align} 
and 
\begin{subequations} 
\begin{align} 
0 &=  r \frac{d \ko_{tt0}}{dr} 
+ \frac{ 5r+2m-4\pi r^3(\mu+p) }{m+4\pi r^3 p} \ko_{tt0} 
- \frac{ r^3 f }{ 2(m + 4\pi r^3 p) } \frac{d \ko_{tr1}}{dr} 
\nonumber \\ & \quad \mbox{} 
+ \frac{ \bigl[ r - 4m + 2\pi r^3 (\mu-3p) \bigr] r}
{m+ 4\pi r^3 p} \ko_{tr1} 
- \frac{5 e^{2\psi}}{r(m+4 \pi r^3 p)} \ko_0 
\nonumber \\ & \quad \mbox{} 
+ \frac{r}{6} \biggl[ 
\frac{ (3-32\pi^2 r^4 p^2) r^2 - 2(3 + 8\pi r^2 p) r m - 2m^2 }
{m + 4\pi r^3 p} \frac{d\omega}{dr} 
- 12 \omega + 12 \biggr]  \frac{d \bq_t}{dr} 
\nonumber \\ & \quad \mbox{} 
- \frac{1}{3(m+4\pi r^3 p)} \biggl\{   
\Bigl[ 2(1 + 10 \pi r^2 p - 16\pi^2 r^4 p^2) r^2 
+ (1 - 16\pi r^2 p) r m - 2m^2 \Bigr] \frac{d\omega}{dr} 
\nonumber \\ & \quad \mbox{} 
+ 2 \Bigl[ 15 r + 6m - 8\pi r^3 (\mu+p) \Bigr] \omega 
- 6 \Bigl[ 5r+2m-4\pi r^3(\mu+p) \Bigr] \biggr\} \bq_t, 
\\ 
0 &=  r \frac{d \ko_{0}}{dr} 
- \frac{ 5r+2m+8\pi r^3 p }{m+4\pi r^3 p} \ko_{0} 
- \frac{ e^{-2\psi} r^5 f}{2(m+4\pi r^3 p)} \frac{d \ko_{tr1}}{dr} 
\nonumber \\ & \quad \mbox{} 
+ \frac{ e^{-2\psi} r^3 \bigl[ r-3m + 2\pi r^3 (\mu-p) \bigr] }
{ m+4\pi r^3 p } \ko_{tr1} 
+ \frac{ e^{-2\psi} r^2 \bigl[ 5r+2m-4\pi r^3(\mu+p) \bigr] } 
{ m+4\pi r^3 p } \ko_{tt0} 
\nonumber \\ & \quad \mbox{} 
+ \frac{ e^{-2\psi} r^4 f (3r+2m+8\pi r^3 p) }{ 6(m+4\pi r^3 p) } 
\frac{d\omega}{dr} \frac{d \bq_t}{dr} 
\nonumber \\ & \quad \mbox{} 
- \frac{2 e^{-2\psi} r^2}{3(m+4\pi r^3 p)} \biggl\{  
\Bigl[ r f (r+m+4\pi r^3 p) \Bigr] \frac{d\omega}{dr}  
\nonumber \\ & \quad \mbox{} 
+ \Bigl[ 15r+6m-8\pi r^3(\mu+p) \Bigr] \omega 
- 3 \Bigl[ 5r+2m - 4\pi r^3(\mu+p) \Bigr] 
\biggr\} \bq_t. 
\end{align}   
\end{subequations} 
The remaining variables are then given by 
\begin{subequations} 
\begin{align} 
\ko_{rr} &= e^{-2\psi} \biggl\{ 
\frac{1}{f} \ko_{tt0} 
- \frac{r^2}{3} \frac{d\omega}{dr} \frac{d\bq_t}{dr} 
+ \frac{2}{3} \biggl[ r \frac{d\omega}{dr} 
+ \frac{ 3(1-\omega) }{f} \biggr] \bq_t \biggr\}, \\ 
\po &= e^{-2\psi} \biggl[
\frac{f}{16\pi} \frac{d \ko_{tr1}}{dr}  
+ \frac{m-2\pi r^3(\mu-p)}{8\pi r^2} \ko_{tr1} 
+ \frac{1}{2} (\mu+p) \ko_{tt0} 
+ \frac{1}{3} (\mu+p)(3-2\omega) \bq_t \biggr],  
\end{align}
\end{subequations}  
and 
\begin{subequations} 
\begin{align} 
\vo_r &= -t\, \frac{ e^{-\psi} \bigl[ 3 - 4\pi r^2(\mu+p) \bigr] }  
{ 4\pi r^2 (\mu+p) } \ko_{tr1}, \\  
\vo &= -t\, \frac{ 3e^{-\psi} }{16\pi r^2 (\mu+p)} \biggl\{ 
r^2 f \frac{d \ko_{tr1}}{dr} 
+ 2 \bigl[ m - 2\pi r^3(\mu-p) \bigr] \ko_{tr1} \biggr\}. 
\end{align} 
\end{subequations} 

As before the interior solutions must be matched with the exterior
solutions of Paper I at $r=R$, and this reveals that $\ko_{tr1}$ must
satisfy the boundary conditions\footnote{The calculation carried out
  in footnote \ref{foot:surface} also reveals that $\ko_{tr1} \propto
  (1-r/R)^{2+n}$ when $r \to R$.}   
\begin{equation} 
\ko_{tr1} = 0 = \frac{d \ko_{tr1}}{dr} \qquad 
\text{at $r=R$}.  
\end{equation} 
These also follow from the requirement that $\Delta p = 0$
at $r= R$, assuming that both $\mu$ and $p$ go to zero at the
surface. The matching also requires $\ko_{tt0}$, $\ko_{rr}$, and
$\ko_0$ to agree at $r=R$ with the values listed in IV of Paper I
(after taking into account the change of notation documented in
footnotes \ref{foot:4} and \ref{foot:5}).  In addition, all these
functions must satisfy regularity conditions at $r=0$. 

As a final exercise we reconsider the interior solution to allow the
fluid to be static. We set $\vo_r = e^{-\psi} \ko_{tr}$ and $\vo = 0$
to ensure that $\delta u^r = 0 = \delta u^A$. As with $\ell = 1$
we find that the conservation equations determine $\po$ and imply that
$\ko_{tr}$ is time-independent. We further assume that the remaining
metric variables are stationary, and verify that the field equations
are consistent with this assumption. We therefore find that a
stationary solution to the $\ell = 3$ interior problem is possible
when the fluid is static, and this conclusion is compatible with the
computations presented in Ref.~\cite{pani-etal:15b}.    

\begin{acknowledgments} 
We are very grateful to John Friedman for a very helpful
conversation. This work was supported by the Natural Sciences 
and Engineering Research Council of Canada.     
\end{acknowledgments}    

\bibliography{../bib/master} 

\end{document}